\documentclass[preprint,aps,prc,tightenlines,floatfix,showpacs,showkeys]{revtex4}

\usepackage{graphicx}% Include figurefiles
\usepackage{dcolumn}
\usepackage{epsfig}
\usepackage{bm}

\begin{document}

\title{Two--neutrino double beta decay  of deformed nuclei within QRPA with realistic interaction}

\author{Mohamed Saleh Yousef}
\affiliation{Institute f\"{u}r Theoretische Physik der Universit\"{a}t
T\"{u}bingen, D-72076 T\"{u}bingen, Germany}
\author{Vadim Rodin}
\email{vadim.rodin@uni-tuebingen.de}
\affiliation{Institute f\"{u}r Theoretische Physik der Universit\"{a}t
T\"{u}bingen, D-72076 T\"{u}bingen, Germany}
\author{Amand Faessler}
%\email{amand.faessler@uni-tuebingen.de}
\affiliation{Institute f\"{u}r Theoretische Physik der Universit\"{a}t
T\"{u}bingen, D-72076 T\"{u}bingen, Germany}
\author{Fedor \v Simkovic}
%\email{fedor.simkovic@fmph.uniba.sk}
\altaffiliation{On  leave of absence from Department of Nuclear
Physics, Comenius University, Mlynsk\'a dolina F1, SK--842 15
Bratislava, Slovakia}
\affiliation{Institute f\"{u}r Theoretische Physik der Universit\"{a}t
T\"{u}bingen, D-72076 T\"{u}bingen, Germany}

\begin{abstract}

A method to implement a realistic nucleon-nucleon residual interaction based on the Brueckner G-matrix (for the Bonn CD force) into the Quasiparticle Random Phase Approximation (QRPA) for deformed nuclei is formulated.
The two-neutrino double beta decay for ground state to ground state transitions 
$^{76}$Ge$ \rightarrow ^{76}$Se and $^{150}$Nd $\rightarrow ^{150}$Sm 
is calculated along with the Gamow-Teller strength distributions. 
The effect of deformation on the observables is studied.

\end{abstract}

\pacs{%PACS Numbers:
21.60.-n, %Nuclear structure models and methods
21.60.Jz, %Nuclear Density Functional Theory and extensions (includes Hartree-Fock and random-phase approximations)
23.40.-s, %beta decay; double beta decay; electron and muon capture
23.40.Hc, %Relation with nuclear matrix elements and nuclear structure 
% 23.40.Bw %Weak-interaction and lepton (including neutrino) aspects (see also 14.60.Pq Neutrino mass and mixing)
}

\keywords{Double beta decay; Nuclear matrix element; Quasiparticle random phase approximation}

\date{\today}

\maketitle

\section{Introduction}

Nuclear double beta decay is a second order weak interaction process which can proceed in two different modes: the two 
neutrino mode $2\nu\beta\beta$, with the emission of two neutrinos and two electrons, and the neutrinoless mode $0\nu\beta\beta$ with emission of two electrons only, without neutrino emission (see, e.g., Refs.~\cite{vogel,hax,fae98,SuCi}). 
Observation of the latter mode that violates the lepton number conservation 
will prove that the neutrino is a massive Majorana particle.
  
Theoretical interpretation 
of electroweak processes in nuclei requires an accurate description of nuclear many-body wave functions. 
The proton-neutron quasiparticle random phase approximation (QRPA), first considered in Ref.~\cite{halb}, 
%to describe $\beta$-decay strength, 
is one of the most reliable nuclear structure methods used for 
describing the structure of the intermediate nuclear states virtually 
excited in double beta decay. 
Important ground state correlations are naturally accounted for within 
the QRPA as well (see, e.g., Refs.~\cite{fae98,SuCi,rod}). 
The method has been shown to be capable of successful describing double beta decay 
provided the particle-particle ($p-p$) residual interaction is included into consideration, along
with the usual particle-hole ($p-h$) one~\cite{2bqrpa}. On the other hand, the calculated nuclear matrix elements for 
double beta decay, both the two-neutrino and neutrinoless modes, have been
proved to sensitively depend on the parameter $g_{pp}$ 
that renormalizes the G-matrix strength in the $p-p$ channel~\cite{fae98,rod}.

As the majority of $\beta\beta$-decaying nuclei are nearly spherical, 
the spherical symmetry has been assumed in all QRPA calculations of 
the $0\nu\beta\beta$-decay matrix elements $M^{0\nu}$ up to now. 
Nowadays, there is a growing interest in the scientific community in double beta decay of
$^{150}$Nd as a very promising candidate for the next generation of experimental 
searches for the neutrinoless double beta decay (SNO+~\cite{SNO+} and SuperNEMO~\cite{SuperNEMO}). 
The nucleus is well known to have one of the largest values of 
the lepton phase space factor (about 33 times %32.72 
larger than that for $^{76}$Ge, see, e.g. Ref.~\cite{vogel}), 
but it is at the same time strongly deformed that provides a great obstacle for reliable theoretical analysis of the corresponding matrix element $M^{0\nu}$. The effect of deformation on  the $0\nu\beta\beta$-decay of $^{150}$Nd to the ground and excited states of $^{150}$Sm was first studied in Ref.~\cite{hirsch} within the pseudo-SU(3) model that is a tractable shell model for deformed nuclei, in which, however, a severely truncated single-particle basis is used.
The matrix element $M^{0\nu}= -1.57$  was found in Ref.~\cite{hirsch} for the 
ground state to ground state (g.s.-to-g.s.) transition $^{150}$Nd$\to ^{150}$Sm, 
substantially smaller than the recent result $M^{0\nu}=4.74$ of Ref.~\cite{rod}. In the latter publication, however, 
the nuclei were considered as spherical. 
Recent result $M^{0\nu}= -1.61$ obtained in Ref.~\cite{hirsch08} within the Projected Hartree-Fock-Bogoliubov approach 
is in a good agreement with the previous one of Ref.~\cite{hirsch}.
Also the double beta decay $^{160}$Gd$\to^{160}$Dy is of interest although the lepton phase space factor is about 6 times smaller than in the case of $^{150}$Nd$\to ^{150}$Sm.

An extension of the pnQRPA method to accommodate the effect of nuclear deformation was first done 
in Ref.~\cite{kru}, where single particle (s.p.) basis was generated in a Nilsson potential. 
Further developments by including Woods-Saxon type 
potentials~\cite{moll}, residual interactions in the particle-particle 
channel~\cite{homma}, selfconsistent deformed Hartree-Fock mean fields 
with consistent residual interactions~\cite{sarr} have followed the article~\cite{kru}.  

Recently, the Gamow-Teller strength distributions and the $2\nu\beta\beta$-decay matrix  
elements have been calculated for deformed nuclei within the QRPA by making use of a phenomenological 
deformed Woods-Saxon potential and the schematic separable 
forces~\cite{gese,deform,al-rod}. It has been found that differences in deformation between initial and final nuclei   
can have a pronounced effect on the $2\nu\beta\beta$-decay half-lives. 
However, using the schematic forces for calculating the amplitudes of the $0\nu\beta\beta$-decay
would immediately raise the problem of how to fix the numerous strength parameters of the forces in different
$J^\pi$ partial channels. Therefore, using realistic interaction with a minimal number of 
renormalization parameters is obviously preferable.
In the present paper the approach used in Refs.~\cite{deform,al-rod} is extended to accommodate realistic effective interaction based on the Brueckner $G$-matrix derived from the nucleon-nucleon Bonn CD force.

The $2\nu\beta\beta$-decay half-lives have been already measured for a dozen of nuclei and 
the corresponding nuclear matrix element $M^{2\nu}_{exp}$ have been extracted~\cite{barab}. 
Theoretical interpretation of these matrix elements  provides a cross 
check of reliability of the calculated nuclear wave functions. 
In recent most detailed spherical QRPA calculations of the  $0\nu\beta\beta$-decay 
matrix elements Ref.~\cite{rod}, the renormalization parameter $g_{pp}$ of the particle-particle residual interaction is fixed in such a way that the experimental half-lives of $2\nu\beta\beta$-decay are correctly reproduced.
Thus, before calculating the matrix elements $M^{0\nu}$ using the QRPA extended to deformed nuclei, 
the corresponding calculation of $M^{2\nu}$  %$2\nu\beta\beta$-decay matrix elements 
has to be done as a test for the modeled nuclear wave functions.

\section{The QRPA in deformed nuclei
%$2\nu\beta\beta$-decay
}

The inverse half-life of the $2\nu\beta\beta$-decay  can be expressed as a product of  an accurately known
phase-space factor $G^{2\nu}$ and the second order Gamow-Teller matrix element $M^{2\nu}_{GT}$ for the g.s.-to-g.s. transition \cite{vogel}:
\begin{equation}
[T^{2\nu}_{1/2}(0^+_{g.s.} \to 0^+_{g.s.}) ]^{-1} =
G^{2\nu} %~(g_A)^4~ 
| M^{2\nu}_{GT}|^2.
\label{l}
\end{equation} 
Contribution from the two successive Fermi transitions to the amplitude of the $2\nu\beta\beta$-decay 
can safely be neglected as it arises from isospin mixing effects (see, e.g., Ref.~\cite{hax}).
%For deformed nuclei 
The double Gamow-Teller matrix element $M^{2\nu}_{GT}$ in Eq.~(\ref{l})
%for g.s.-to-g.s. $2\nu\beta\beta$-decay transition  acquires the form
can be written in the following form:
\begin{equation}
M^{2\nu}_{GT}=\sum\limits_{m}
\frac{\langle 0^+_f\| \beta^- \| m\rangle \langle m\| \beta^- \| 0^+_i\rangle}{\bar\omega_m},
\label{2.0}
\end{equation}
where the index $i (f)$ refers to the initial (final) nuclei, the sum runs over all $| m\rangle=
%\equiv
|1^+\rangle$ states of the intermediate odd-odd nucleus, $\beta^-=\sum_a \mbox{\boldmath $\sigma$}_a \tau^-_a$ is the Gamow-Teller transition operator, 
$\bar\omega_m=E_m-(E_{0_i}+E_{0_f})/2=(\omega_{m(i)}+{\omega}_{m(f)})/2$ with
$\omega_{m(i)}=E_m-E_{0_i}$ ($\omega_{m(f)}=E_m-E_{0_f}$) representing
the excitation energy of the $m$'th state relative to the g.s. of the initial (final) nucleus.

To take into account the effect of deformation, wave functions $|1^+\rangle$ of the intermediate states 
in the laboratory frame which have a projection $M$ of the total angular momentum onto the axis $z$ can be represented in terms of wave functions in the intrinsic frame:
\begin{eqnarray} 
|1M(K),m\rangle &=& \sqrt{\frac{3}{16 \pi^2}}
[ {\cal D}^1_{M K}(\phi,\theta,\psi) Q_{m,K}^\dagger + 
(-1)^{1+K} {\cal D}^1_{M -K}(\phi,\theta,\psi) 
Q_{m,-K}^\dagger ] |0^+_{g.s.}\rangle ~~~(K=\pm1),
\nonumber \\
|1M(K),m \rangle &=& \sqrt{\frac{3}{8 \pi^2}}
{\cal D}^1_{M K}(\phi,\theta,\psi) Q_{m,K}^\dagger |0^+_{g.s.}\rangle
~~~(K=0).
\label{eq12}
\end{eqnarray}
Here, the correlated QRPA ground state in the intrinsic frame is denoted as $|0^+_{g.s.}\rangle$, 
the intrinsic excitations are generated by the QRPA phonon creation operator $Q_{m,K}^\dagger$,
%, and the states $|1M(K),m\rangle$  %the proton-neutron QRPA phonon   wave functions 
%for Gamow-Teller excitations in the intermediate  nucleus have the form 
and $K$ is the projection of the total angular momentum 
onto the nuclear symmetry axis (the only projection which is conserved in strongly deformed nuclei). 

These adiabatic Bohr-Mottelson--type wave functions provide an approximation which is valid
for large deformations. Thus, the most interesting nuclei in question, $^{150}$Nd and $^{150}$Sm, which indeed are strongly deformed, are well justified to be treated in such an approximation.
The adiabatic, or the strong coupling (see, e.g., the monograph~\cite{RingSchuck80}), approach fails, however, for small deformations since the Coriolis force gets large and mixes states with different $K$. 
Nuclei $^{76}$Ge and $^{76}$Se have rather small deformations and the so-called weak coupling, or no alignment, limit~\cite{RingSchuck80} seems to be more suitable. In this limit the Coriolis force becomes so strong that the angular momenta of the valence nucleons get completely decoupled from the orientation of the core. 
%and $K$ is not a good quantum number anymore. 
Such a case would deserve a detailed study that is postponed to a future publication and the adiabatic approach to description of excited states of deformed nuclei is adopted in the present first application of the QRPA with a realistic residual interaction. Nevertheless, one might already anticipate without calculations that in the weak coupling limit the calculated observables should reveal smaller deviations from the ones obtained in the spherical limit than those calculated in the strong coupling limit of the present work. In this connection it is worth to note that spherical QRPA results can exactly be reproduced in the present calculation by letting deformation vanish, in spite of the formal inapplicability of the strong-coupling ansatz for the wave function in this limit.

%In this work we do not take into account this effect as we deal always with quantities which are sums over different $K$.

The 
%intrinsic states are generated by 
QRPA phonon creation operator acting on the ground-state wave function is given as: 
\begin{equation}
Q_{m,K}^\dagger = \sum_{pn} X^{m}_{pn, K} A^\dagger _{pn, K} - Y^{m}_{pn, K} \bar{A}_{pn,K}.
\label{3}
\end{equation}
Here, $A^\dagger _{pn,K}=a^\dagger_{p}{a}^{\dagger}_{\bar{n}}$ and $\bar{A}_{pn,K}={a}_{\bar{p}}{a}_{n}$ 
are the two-quasiparticle creation and annihilation operators, 
respectively, with the bar denoting the time-reversal operation.
The quasiparticle pairs $p\bar n$ are defined by the selection rules $\Omega_p -\Omega_n = K$ 
and $\pi_p\pi_n=1$, where $\pi_\tau$ is the single-particle (s.p.) parity and  $\Omega_\tau$ is the projection 
of the total s.p. angular momentum on the nuclear symmetry axis ($\tau = p,n$).
The s.p. states $|p\rangle$ and $|n\rangle$ 
of protons and neutrons are calculated by solving the Schr\"odinger equation with the deformed axially symmetric 
Woods-Saxon potential \cite{Damgaard,Nojarov}. In the cylindrical coordinates the  
deformed Woods-Saxon s.p. wave functions $|\tau \Omega_\tau\rangle$ with $\Omega_\tau>0$ are 
%calculated as 
decomposed over the deformed harmonic oscillator 
s.p. wave functions (with the principal quantum numbers $(N n_{z} \Lambda)$) and the spin wave functions 
$|\Sigma=\pm \frac12\rangle$:
%are expressed as:
 \begin{eqnarray}
|\tau \Omega_\tau\rangle &=& \sum_{N n_z \Sigma} b_{N n_z \Sigma} |N n_z \Lambda_\tau=\Omega_\tau-\Sigma\rangle |\Sigma\rangle,
%\nonumber \\&&~~~~~+ b^{(-)}_{N n_z}  |Nn_z\Lambda_\tau=\Omega_\tau+\frac12\rangle|\downarrow\rangle]
\label{tau}
\end{eqnarray}
where $N=n_\perp+n_z$ ($n_\perp=2n_\rho+|\Lambda|$), $n_z$ and $n_\rho$ are the number of nodes
of the basis functions in the $z$- and $\rho$-directions,
respectively; $\Lambda =  \Omega - \Sigma$ and $\Sigma$ are the
projections of the orbital and spin angular momentum onto the symmetry axis $z$.
For the s.p. states with the negative projection $\Omega_\tau=-|\Omega_\tau|$, 
that are degenerate in energy with $\Omega_\tau=|\Omega_\tau|$,
the time-reversed version of Eq.~(\ref{tau}) is used as a definition (see also Ref.~\cite{deform}).
The states $(\tau,\bar \tau)$ comprise the whole single-particle model space.

The deformed harmonic oscillator wave functions $|N n_z \Lambda\rangle$  
can be further decomposed over the spherical harmonic oscillator ones   
$|n_rl\Lambda\rangle$ by calculating the corresponding spatial  
overlap integrals $A^{n_rl}_{N n_z \Lambda} =\langle n_r l \Lambda|N n_{z} \Lambda\rangle$  
%$N'=2n_r+l$ 
($n_r$ is the radial quantum number,  $l$ and $\Lambda$ are the orbital angular momentum and its projection onto $z$-axes, respectively), see Appendix for the details.   
Thereby, the wave function (\ref{tau}) can be reexpressed as   
\begin{eqnarray} 
|\tau \Omega_\tau\rangle&=& \sum_{\eta}B^{\tau}_{\eta}|\eta\Omega_\tau\rangle , 
\label{4} 
\end{eqnarray} 
where $|\eta\Omega_\tau\rangle=\sum\limits_{\Sigma} C^{j\Omega_\tau}_{l~ \Omega_\tau-\Sigma~\frac12~\Sigma} 
|n_r l \Lambda=\Omega_\tau-\Sigma\rangle |\Sigma\rangle$ is the spherical harmonic oscillator wave function in the $j$-coupled scheme ($\eta=(n_rlj)$), and  
$B^{\tau}_{\eta}= \sum\limits_{\Sigma}C^{j\Omega_{\tau}}_{l ~\Omega_{\tau}-\Sigma~\frac12~\Sigma}\, A^{n_r l}_{N n_z \Omega_{\tau}-\Sigma}\, b_{N n_z \Sigma}$, with $C^{j\Omega_{\tau}}_{l ~\Omega_{\tau}-\Sigma~\frac12~\Sigma}$ being the Clebsch-Gordan coefficient. 

The  QRPA equations:
\begin{eqnarray}
\left( \matrix{ {\cal A}(K) & {\cal B}(K) \cr
-{\cal B}(K) & -{\cal A}(K) }\right) 
~\left( \matrix{ X^m_K \cr Y^m_K} \right)~ = ~
\omega_{K,m}
%~\left( \matrix{ 1 & 0 \cr 0 & -1 }\right) 
~\left( \matrix{ X^m_K \cr Y^m_K} \right),
\label{5}
\end{eqnarray}
with realistic residual interaction are solved to get the forward $X^m_{i K}$,  
backward $Y^m_{i K}$ amplitudes and the excitation energies $\omega^{m_i}_K $ and $ \omega^{m_f}_K$  of  the $m$-th 
$K^+$ ($ K=0,\pm 1$) %$1^+$ 
state in the intermediate nucleus. The matrix $\cal A$ and $\cal B$ are defined by
\begin{eqnarray}
{{\cal A}_{p n,{p'}{n'}}}(K)&=&{\delta}_{p n,{p'}{n'}}(E_p+E_n)+g_{pp}(u_p u_n u_{p'} u_{n'}+ v_p v_n v_{p'} v_{n'}) 
V_{p \bar{n}p'\bar{n'}}\nonumber\\
&&~~~~~~~~~~~~~~~~~~~~~~-g_{ph}(u_p v_n u_{p'} v_{n'}+ v_p u_n v_{p'} u_{n'})
V_{p n'p'n}\nonumber\\
{{\cal B}_{p n,{p'}{n'}}}(K)&=&-g_{pp}(u_p u_n v_{p'} v_{n'}+ v_p v_n u_{p'} u_{n'}) 
V_{p \bar{n}p'\bar{n'}}\nonumber\\
&&-g_{ph}(u_p v_n v_{p'} v_{n'}+ v_p u_n u_{p'} v_{n'})
V_{p n'p'n}
%\langle p \rho_p \bar{n}\rho_n| G |p'\rho_{p'} \bar{n'}\rho_{n'}\rangle
\label{6}
\end{eqnarray}  
where $E_p+E_n$ are the two-quasiparticle excitation energies, 
$V_{p n,{p'}{n'}}$ and $V_{p \bar {n},{p'}\bar {n'}}$ are the $p-h$ and $p-p$
matrix elements of the residual nucleon-nucleon interaction $V$, respectively,
$u_\tau$ and $v_\tau$ are the coefficients of the Bogoliubov transformation.
The amplitudes of $\beta^-$ and $\beta^+$ transitions  from the $0^+$
g.s. of initial and final nuclei to a one-phonon $K^+$ state in the intermediate nucleus 
%in terms of the single particle matrix elements of the spin-isospin operator$\tau^+ \Sigma_K$ 
are given in the intrinsic system by:
\begin{eqnarray}
\langle K^+,m| {\beta}^-_K |0^+_{g.s.}\rangle &=& 
\sum_{pn} \langle p| \sigma_K| n \rangle \left[  u_p v_n X^m_{pn, K} + v_p u_n Y^m_{pn, K} \right],
\nonumber \\
\langle K^+,m| {\beta}^+_K |0^+_{g.s.}\rangle &=& 
\sum_{pn} \langle p| \sigma_K| n \rangle \left[ u_p v_n  Y^m_{pn, K} + v_p u_n  X^m_{pn, K} \right].
%\nonumber\\
\label{7}
\end{eqnarray}

The matrix element $M^{2\nu}_{GT}$ of Eq.~(\ref{2.0}) is given within the QRPA in the intrinsic system by the following
expression:
\begin{equation}
M^{2\nu}_{GT}= \sum_{K=0,\pm 1} \sum_{{m_i m_f}}
\frac{\langle 0^+_f| \bar\beta^-_{K} | K^+,m_f\rangle\langle K^+,m_f|K^+,m_i\rangle
\langle K^+,m_i| \beta^-_K | 0^+_i\rangle}{\bar\omega_{K,m_im_f}}.
\label{2}
\end{equation}
Along with the usual approximation of the energy denominator in Eq.~(\ref{2}) as ${\bar\omega}_{K,m_im_f}=(\omega_{K,m_f} + \omega_{K,m_i})/2$ (see, e.g., Refs.~\cite{deform,al-rod}; we will later refer to this case as ``case II"), we also use in this work another prescription in which the whole calculated QRPA energy spectrum is shifted in such a way as to have the first calculated $1^+$ state exactly at the corresponding experimental energy (case I). In this case the energy denominator in Eq.~(\ref{2}) acquires the form ${\bar\omega_{K,m_im_f}}=(\omega_{K,m_f} - \omega_{K,1_f} + \omega_{K,m_i}  - \omega_{K,1_i})/2+\bar\omega_{1^+_1}$, with $\bar\omega_{1^+_1}$ being the experimental excitation energy of the first $1^+$ state measured from the mean g.s. energy $(E_{0_i}+E_{0_f})/2$. All the calculated strength functions in this work are represented according to the case I, as well. 

The two sets of intermediate nuclear states generated from the
initial and final g.s. do not come out identical within the
%considered approximation scheme
QRPA. Therefore, the overlap factor
of these states is introduced in Eq.~(\ref{2}) \cite{wies,pan88} as follows:
\begin{equation}
\langle K^+,m_f|K^+,m_i\rangle =
\sum_{l_i l_f}
~[X^{m_f}_{l_f K}X^{m_i}_{l_i K}-Y^{m_f}_{l_f K}Y^{m_i}_{l_i K}] 
\, {\cal R}_{l_f l_i}
\, \langle BCS_f|BCS_i\rangle.
\label{8}
\end{equation}
The factor ${\cal R}_{l_f l_i}$, which includes the overlaps of single particle 
wave functions of the initial and final nuclei is given  by:
\begin{eqnarray}
{\cal R}_{ll'}&=& \langle p \rho_p |p' {\rho}_{p'} \rangle(u^{(i)}_p u^{(f)}_{p'}+v^{(i)}_p v^{(f)}_{p'}) 
 \langle n \rho_n |n' {\rho}_{n'} \rangle(u^{(i)}_n u^{(f)}_{n'}+v^{(i)}_n v^{(f)}_{n'}),
\label{9}
\end{eqnarray} 
and the last term $\langle BCS_f|BCS_i\rangle$ in Eq.~(\ref{8}) corresponds to the overlap factor of the initial and final BCS vacua in the form given in Ref.~\cite{deform}.

As a residual two-body interaction we use the nuclear Brueckner G-matrix, that is a solution of the Bethe-Goldstone equation, derived from the Bonn-CD one boson exchange potential, 
as used also in the spherical calculations of Ref.~\cite{rod}. 
The G-matrix elements are originally calculated with respect to a spherical harmonic oscillator s.p. basis.
By using the decomposition of the deformed s.p. wave function in Eq.~(\ref{4}), 
the two-body deformed wave function %with total angular momentum and parity $J^\pi$ 
can be represented as:
\begin{eqnarray}
|p \bar{n}\rangle&=& \sum_{\eta_p\eta_n J}
%B_{p\eta_p}B_{n\eta_n}\sum_{J}(-1)^{j_n-\Omega_{n}}C^{JK}_{j_p\Omega_{p} j_n-\Omega_{n} }
F^{JK}_{p\eta_p n\eta_n}|\eta_p \eta_n,J K\rangle,
\label{decomp}
\end{eqnarray}
where 
$|\eta_p \eta_n,J K\rangle=\sum_{J}C^{JK}_{j_p\Omega_{p} j_n\Omega_{n} }|\eta_p \Omega_{p}\rangle|\eta_n \Omega_{n}\rangle$,
and  
$F^{JK}_{p\eta_p n\eta_n}= B^p_{\eta_p}B^n_{\eta_n}(-1)^{j_n-\Omega_{n}}C^{JK}_{j_p\Omega_{p} j_n-\Omega_{n}}$ 
is defined for the sake of simplicity
($(-1)^{j_n-\Omega_{n}}$ is the phase arising from the time-reversed states $|\bar{n}\rangle$).
The particle-particle $V_{p \bar {n},~{p'}\bar {n'}}$ and particle-hole $V_{p n',~p'n}$ 
interaction matrix elements in the representation (\ref{6}) 
for the QRPA matrices ${\cal A,\ B}$ (\ref{5}) in the deformed Woods-Saxon 
single-particle basis can then be given in terms of the spherical G-matrix elements as follows:
\begin{eqnarray}
%\langle p \bar{n}|V|p' \bar{n'}\rangle 
V_{p \bar {n},~{p'}\bar {n'}}&= -&
2\sum_{J}\sum_{{\eta}_{p}{\eta}_{n}} \sum_{{\eta}_{p'}{\eta}_{n'}}
%\sum_{\begin{array}{l}\ {\eta}_{p}{\eta}_{n}\\\ {\eta}_{p'}{\eta}_{n'}\end{array}}
F^{JK}_{p\eta_p n\eta_n}F^{JK}_{p'\eta_{p'} {n'}\eta_{n'}}
G(\eta_p\eta_n\eta_{p'}\eta_{n'},J),\\
V_{p n',~p'n}&=&
2\sum_{J}\sum_{{\eta}_{p}{\eta}_{n}} \sum_{{\eta}_{p'}{\eta}_{n'}}
F^{JK'_{pn'}}_{p\eta_p {\bar n}'\eta_{n'}}
F^{JK'_{pn'}}_{p'\eta_{p'} \bar{n}\eta_{n}} G(\eta_p\eta_{n'}\eta_{p'}\eta_{n},J),
\label{10}
\end{eqnarray}
where $K'_{pn'}=\Omega_p+\Omega_{n'}=\Omega_{p'}+\Omega_n$.
The matrix elements of $\sigma_K$ in Eq.~(\ref{7}) can be written as
$\langle p| \sigma_K|n\rangle= \sum_{\eta_p,\eta_n}
F^{1K}_{p\eta_p n\eta_n}\langle\eta_p \|\sigma \| \eta_n\rangle /\sqrt{3}$.

\section{results}

The Gamow-Teller strength distributions and the $2\nu\beta\beta$-decay amplitudes for the g.s.-to-g.s.
transitions are calculated for the nuclear systems with
$A$ = 76  ($^{76}$Ge $\rightarrow ^{76}$Se)
%, $A$ = 100 ($^{100}$Mo $\rightarrow ^{100}$Ru), 
and $A$ = 150 ($^{150}$Nd $\rightarrow ^{150}$Sm). 
The single-particle Schr\"odinger equation with the Hamiltonian of 
a deformed Woods-Saxon mean field~\cite{Damgaard,Nojarov}  is solved on the basis 
of a axially-deformed %symmetric 
harmonic oscillator, see Eq.~(\ref{tau}).
The s.p. basis corresponding in the spherical limit to full (2--4)$\hbar\omega$
major oscillator shells for the nuclei with $A$ = 76 
and (4--6)$\hbar\omega$ for $A$ = 150 are used.
Decomposition over the states within seven major spherical harmonic oscillator shells is done in Eq.~(\ref{decomp}).
Only quadrupole deformation is taken into account in the calculations. The deformation parameter $\beta_2$ is obtained 
as $\beta_2= \sqrt{\frac{\pi}{5}}\frac{Q_p}{Z r^{2}_{c}}$ ($r_c $ is the charge rms radius) by using the empirical intrinsic quadrupole moments $Q_p$ which are derived from the laboratory quadrupole moments measured by the Coulomb excitation reorientation technique~\cite{ragha}. The corresponding values of 
$\beta_2$ are listed in Table~\ref{table.1}. 
The spherical limit, i.e. $\beta_2=0$, is considered in all cases as well.
Because of rather large experimental errors $\beta_2=0.37 \pm 0.09$ and 
$\beta_2=0.23 \pm 0.03$~\cite{ragha} for $^{150}$Nd and $^{150}$Sm, respectively,  
we also adopt for these nuclei the respective calculated values from Ref.~\cite{moeller} which seem to fit better 
the rotational bands in these nuclei.

First, the BCS equations are solved self-consistently to obtain the occupation
amplitudes $u_\tau$ and $v_\tau$, gap parameter $\Delta_\tau$ and the chemical potentials $\lambda_p$ 
and $\lambda_n$~\cite{moh}. The renormalizing strengths $g{^{p}_{pair}} $ and $ g^{n}_{pair}$ of the proton and neutron pairing interactions are determined to reproduce the experimental pairing energies through 
a symmetric five term formula~\cite{Audi}.

For calculating the QRPA energies and wave functions 
one has to fix the particle-hole $g_{ph}$ and particle--particle
$g_{pp}$ renormalization factors of the residual interaction in Eq.~(\ref{6}). 
An appropriate value of $g_{ph}$ can be determined 
by reproducing the  experimental  position of the Gamow-Teller giant resonance (GTR) in the intermediate nucleus, 
 %\cite{homa,taig}
whereas the parameter $g_{pp}$ can be determined from fitting the experimental
value $M^{2\nu-exp}_{GT}$. The experimental energy position of the GTR  relative to the energy of the first excited $1^+$ state in $^{76}$Ge can be reproduced with $g_{ph}=1.15$ (the prescription of case I is used, see explanations 
after Eq.~(\ref{2})). Since there is no experimental information on the GTR energy for $^{150}$Nd, we use for this nucleus the same $g_{ph}=1.15$. %as obtained for $^{76}$Ge. 
Two sets of parameters $g_{pp}$ obtained from fitting $M^{2\nu-exp}_{GT}$ in two calculations of $M^{2\nu}_{GT}$ (case I and case II) are listed in Table \ref{table.1}. The difference between $g_{pp}$ fitted in case I and case II is usually quite small. To compare with the previous QRPA results, 
we have performed calculations also with the separable $p-h$ and $p-p$ interactions in a way similar to what was done in Refs.~\cite{deform,al-rod}. The coefficient in the $A$-dependence of the $p-h$ strength parameter $\chi=3.73/A^{0.7}$ MeV 
is fitted to reproduce the GTR energy in $^{76}$Ge, and is used then in the calculations for $^{150}$Nd (the form of the 
$A$-dependence is taken from Ref.~~\cite{homma}).  
The corresponding fitted values of the strength parameters $\kappa$ 
of the separable $p-p$ interaction are also listed in Table \ref{table.1}.
%(they differ slightly from those obtained in Ref.~\cite{deform,al-rod}). 
All the calculations in this work are done with the mean field spin-orbit coupling constant increased by factor 1.2 
as compared with the one used in Refs.~\cite{deform,al-rod}, %Ref.~\cite{Noj2}, 
that gives a better correspondence with the parametrization of 
the spherical Woods-Saxon mean field used in Ref.~\cite{rod}.
Therefore, we get slightly different fitted values of $\chi$ and $\kappa$ as compared with Ref.~\cite{deform,al-rod}.

\begin{table}[h]
% \hspace{0.033\textwidth}
% \begin{minipage}{0.4\textwidth}
%\begin{center}
\centering
\caption{The values of the deformation parameter $\beta_2$ for initial (final) nuclei adopted in the calculations along
with the fitted values of the $p-p$ strength parameters $g_{pp}$ (for the realistic Bonn-CD force) and $\kappa$ (for a phenomenological separable force) for two ways of calculations of $M^{2\nu}_{GT}$, (I) and (II). The $p-h$ strength parameters $g_{ph}=1.15$ and $\chi=3.73/A^{0.7}$ MeV are fixed as explained in the text. In the last column the calculated (with the Bonn-CD force) values of the Ikeda sum rule (in \% of $3(N-Z)$) are given for the initial (final) nucleus.}
\label{table.1}
\begin{tabular}{|l|l|c|c|c|c|c|c|c|}
	\hline	
nucleus    &\ \ \ \  $\beta_2$ & $g_{pp}$ (I) & $g_{pp}$ (II)& $\kappa$ (I), MeV & $\kappa$ (II), MeV &ISR(\%)\\ 
	\hline
$^{76}$Ge ($^{76}$Se) & 0.0\ \ (0.0)              & 0.94 & 0.91 & 0.087 &  0.083 & 96.8\ \ (98.2)\\
                       & 0.10 (0.16) \cite{ragha} & 0.99 & 0.97 & 0.091 &  0.088 & 96.1\ \ (96.8)\\
%                       & 0.10 (0.16) \cite{ragha} & 0.993 & 0.970 & 0.091 &  0.088 & 96.1\ \ (96.8)\\
\hline
$^{150}$Nd ($^{150}$Sm) & 0.0\ \  (0.0)        & 1.11 & 1.11 & 0.051 & 0.050  & 94.8\ \ (95.9)\\
                   & 0.37 (0.23) \cite{ragha}  & 0.78 & 0.65 & 0.033 & 0.005  & 94.1\ \ (95.8)\\
                   & 0.24 (0.21) \cite{moeller}& 1.35 & 1.32 & 0.053 & 0.052  & 95.4\ \ (96.2)\\
%$^{150}$Nd ($^{150}$Sm) & 0.0\ \  (0.0)        & 1.11 & 1.11  & 0.051 & 0.0495  & 94.8\ \ (95.9)\\
%                   & 0.37 (0.23) \cite{ragha}  & 0.784 & 0.646 & 0.0325 & 0.0045  & 94.1\ \ (95.8)\\
%                   & 0.24 (0.21) \cite{moeller}& 1.350 & 1.32  & 0.0534  & 0.052   & 95.4\ \ (96.2)\\
\hline 
\hline
\end{tabular}
%\end{center}
%\end{minipage}
 %\hspace{0.033\textwidth}
\end{table}

The calculation of single $\beta^-$ and $\beta^+$ decay branches for parent and daughter nuclei is the starting point for  the calculation of the $2\nu\beta\beta$-decay amplitudes.
The calculated Gamow-Teller strength distributions for all the nuclei in question 
are shown in Figs.~1,2 as functions of the excitation energy in the intermediate (for the $2\nu\beta\beta$ decay) nuclei. The representation is according to the convention of the case I, i.e. the entire calculated GT spectrum is shifted in order to fit the experimental energy of the first $1^+$ state. %(44 keV in $^{76}$As)
In order to facilitate comparison among various calculations the 
Gamow-Teller distributions are smoothed with a Gaussian of width 1~MeV, so the original discrete spectrum of $B(GT)$ values is transformed into a continuous one of the strength function $S{(GT)}$. 
The relevant strength  distributions for the double beta decay are the $S({GT}^-)$ for the  parent nucleus (two upper panels) and the $S({GT}^+)$ distribution for the daughter nucleus (two lower panels). 
In the left panels labeled as ``Realistic" the results %of the present work 
obtained by using a realistic nucleon-nucleon interaction (Bonn CD force) are shown whereas 
in the right ones labeled as ``Separable" the results obtained by using a separable interaction of Ref.~\cite{al-rod}. In both types of calculations the corresponding strengths of the $p-h$ interaction $g_{ph}$ and $\chi$ are fixed by fitting the experimental energy of the Gamow-Teller resonance in $^{76}$Ge as described in the preceding paragraph. 
The thick solid and the dashed lines represent the results obtained in the spherical limit $\beta_2=0$ %in the calculation 
and with the realistic deformation, respectively. 
The thin solid line in Fig.~1 represents the Gaussian-smeared experimental $B(GT)$ values for $^{76}$Ge taken from Ref.~\cite{madey}.

In the case of $\beta^-$ distribution, one observes that the position of the Gamow-Teller resonance is not sensitive to the effect of deformation.
%but there are significant differences between both, $\beta^-$ and $\beta^+$, cases in the strength distributions.
In the case of $\beta^+$ distribution, the effect of deformation is more apparent in the strength distribution than that in the case of the $\beta^-$ one.
Comparing the results obtained with the realistic and schematic residual interaction one can see some
marked differences in the $\beta^+$ and the low-energy part of the $\beta^-$ strength distributions. 
The Ikeda sum rule is underestimated by a small amount about (3--5)\% in the calculations 
(see last column of Table I; to get 100 \% one would need to have the whole s.p. basis, as it is done, for instance, within the continuum-QRPA, see e.g. Ref.~\cite{rod08}). 
It also means that the chosen s.p. model spaces (see the beginning of this section) are large enough.

Fig. 3 illustrates the evolution of the strength functions $S({GT}^-)$ in $^{76}$Ge and $^{150}$Nd 
with respect to increase of $g_{pp}$ in the spherical limit ($\beta_2=0$). 
One can see in the figure that all the GT peaks get shifted to smaller energies as $g_{pp}$ increases. In addition, the low-lying GT states become more collective and, correspondingly, the low-energy part of the GT strength gets markedly larger at the expense of a decrease in the GTR strength. It can also be seen that the QRPA calculations with the realistic $g_{pp}$ as listed in Table I are still quite far from the collapse of the QRPA.
One may argue that the results show too enhanced %collective 
low-energy part of the GT spectrum as compared with the experiment. 
It might have to do with neglect of quenching in the present calculation. 
For $g_A<1.25$ the corresponding experimental value of $M^{2\nu-exp}_{GT}$ gets larger and a smaller $g_{pp}$ would be needed to fit it. 
This effect deserves a separate detailed study, similar to the one performed in Ref.~\cite{lisi07},
that is out of scope of the present work.

As shown in Ref.~\cite{deform,al-rod}, deformation introduces a mechanism of suppression of 
the $M^{2\nu}_{GT}$ matrix element that works even for the same initial and final deformations. 
%that is different from the $g_{pp}$ one. 
The reduction gets even stronger when initial and final deformations differ from each other, and is mainly due to a decrease of the BCS overlap~(\ref{8}). The values of the overlap calculated with the realistic pairing interaction are very close to those obtained by using the constant pairing gap as in Refs.~\cite{deform,al-rod} 
that is illustrated in Fig.~4 for the transition $^{76}$Ge$\to^{76}$Se.
In addition, there is the well-known reduction of 
the $M^{2\nu}_{GT}$ with increasing renormalization parameter $g_{pp}$ of the $p-p$ interaction \cite{rod,2bqrpa}.
It is interesting to study all these suppression mechanisms within the QRPA with the realistic residual interaction 
that is main thrust of this work.

The calculated matrix elements $M^{2\nu}_{GT}$ are shown in Figs.~5,6 as functions of
the parameter $g_{pp}$ and for different deformations of initial and final nuclei.
The parameters $\beta_2$, $g_{ph}$ and $\chi$ are used the same as in the study of the Gamow-Teller distributions.
%(see Table I). 
Again, in the left panels labeled as 
"Realistic" the results of present work using the realistic nucleon-nucleon interaction (Bonn CD force) are shown whereas 
in the right ones labeled as "Separable" the results using separable interaction are presented.
The upper panels represent the results obtained with the shifted calculated QRPA spectrum (case I, see explanation after Eq.~(\ref{2})) and in the lower ones
the results obtained with the usual unshifted QRPA spectrum (case II) are shown.
The solid lines in Figs.~5,6 represent the  matrix elements $M^{2\nu}_{GT}$ calculated in the spherical limit 
while the dashed ones (and dot-dashed in Fig.~6) represent the $M^{2\nu}_{GT}$ calculated for realistic deformations.
The dotted horizontal line corresponds to the corresponding experimental values $M^{2\nu-exp}_{\rm GT}$ 
obtained in Ref.~\cite{barab} by using  unquenched value of the axial-vector coupling constant $g_A=1.25$. 
The points A and B (and C for $^{150}$Nd$\to^{150}$Sm) in each panel specify the values of the $p-p$ interaction for which the value of $M^{2\nu-exp}_{GT}$ is reproduced for spherical and deformed cases, respectively (the corresponding values of $g_{pp}$ and $\kappa$ are listed in Table I). 
%The corresponding points in the case of the separable interaction are C and D.
The calculated $M^{2\nu}_{GT}$ decreases faster in the spherical case than 
in the  deformed one as $g_{pp}$ increases, and the fitted value of $g_{pp}$ 
is usually larger in the case of deformation than that for the spherical limit.
Another interesting feature is that for non-zero deformation the calculated $M^{2\nu}_{GT}$ is smaller at $g_{pp}=0$
than those in the spherical limit, in agreement with the previous results of Refs.~\cite{deform,al-rod}. 
This suppression already holds when initial and final nuclei are equally deformed and gets stronger
with increasing difference in deformation between initial and final nuclei
(the BCS overlap plays an important role in the reduction of $M^{2\nu}_{GT}$, as discussed in Refs.~\cite{deform,al-rod}). 
It is also worth to mention that the calculated $M^{2\nu}_{GT}$ for $^{150}$Nd$\to^{150}$Sm transition (Fig.~6) 
comes out rather small when the experimental deformation parameters from Ref.~\cite{ragha} are used, while by using the deformation parameters of Ref.~\cite{moeller} the corresponding $M^{2\nu-exp}_{GT}$ can be fitted for a reasonable value of $g_{pp}$.
Also, the value of $M^{2\nu}_{GT}$ calculated with the shifted QRPA spectrum, though a little bit larger at $g_{pp}\ll 1$ 
than the unshifted ones, fit the corresponding experimental values at almost the same $g_{pp}$ (see Table I).

\section{Conclusions}

In the present work the two-neutrino double beta decay (g.s.-to-g.s. transitions) 
and relevant Gamow-Teller strength distributions are calculated within the QRPA 
for the nuclear systems 
$^{76}$Ge$ \rightarrow ^{76}$Se %, $^{100}$Mo$ \rightarrow ^{100}$Ru, 
and $^{150}$Nd$ \rightarrow ^{150}$Sm 
with taking into account effects of nuclear deformation.
For the first time a realistic residual two-body interaction based on the Brueckner G-matrix (for Bonn CD force)
is implemented in deformed calculations. The G-matrix elements in the deformed Wood-Saxon basis 
are calculated by expanding the deformed single-particle wave functions over the spherical harmonic oscillator ones.
The effect of deformation on the observables is studied within the framework. The suppression of the calculated $M^{2\nu}_{GT}$, due to both non-zero deformation and the interaction in the particle-particle channel, is observed 
in accordance with previous calculations with separable forces ~\cite{deform,al-rod}. 
The present work realizes the first important step towards the QRPA calculations of 
the neutrinoless double beta decay of deformed nuclei like $^{150}$Nd with a realistic nucleon-nucleon interaction.

\acknowledgments

The authors acknowledge support of the Deutsche Forschungsgemeinschaft within both the SFB TR27 "Neutrinos and Beyond" and Graduiertenkolleg GRK683, and the EU ILIAS project under the contract RII3-CT-2004-506222.

\appendix\section{Calculation of the overlap integrals}

In this Appendix we describe how the deformed harmonic oscillator wave functions 
can be decomposed over the spherical ones by calculating the corresponding spatial overlap integrals.

The normalized wave equation of the three-dimensional axially-deformed 
harmonic oscillator in cylindrical coordinates $(\rho, z, \phi)$ are given by a product of three functions:
\begin{eqnarray}
|N n_z \Lambda\rangle =\psi^{|\Lambda|}_{n_\rho}(\rho)~\psi_{n_z}(z)~\frac{e^{i\Lambda \phi}}{\sqrt{2\pi}},
%\psi_{\Lambda}(\phi)
\end{eqnarray} 
where  $\Lambda$  the projection  of the  orbital angular momentum on $z$ (symmetry) axis,
$N$ is the principal quantum number which is defined as $N=n_z+2n_\rho+|\Lambda|$.
The radial function $\psi ^{|\Lambda|}_{n_\rho}$ is usually written in terms  of a dimensionless coordinate $\eta$ as 
\begin{equation}
\psi ^{|\Lambda|}_{n_\rho}(\rho)= C^{|\Lambda |}_{n_\rho}\eta^\frac{|\Lambda |}{2}
e^{-\frac{\eta}{2}}L^{|\Lambda|}_{n_\rho}(\eta)
\end{equation} 
with $\eta= \frac{\rho^2}{b_\perp^{2}}$, where $ b_\perp=\sqrt{\frac{\hbar}{m\omega_\perp}}$ is 
the oscillator length for the motion perpendicular to the z-axis, 
$C^{|\Lambda |}_{n_\rho}=  \bigg(\frac{2n_\rho !}{(n_{\rho} +|\Lambda |)!\, b_\perp^2} \bigg)^\frac{1}{2}$
is a normalization factor and $L^{|\Lambda|}_{n_\rho}(\eta)$  are the associated Laguerre polynomials.                      
The z-dependent function $\psi_{n_z}$ is similarly written in terms of a dimensionless variable $\xi$ as 
\begin{equation}
\psi_{n_z}(z)=C_{n_z}e^{-\frac{\xi^2}{2}}H_{n_z}(\xi)
\end{equation} 
with $\xi= \frac{z}{b_z}$, here $b_z$ is oscillator length in the direction of the 
z-axis, $C_{n_z}=(\sqrt{\pi}2^{n_z} n_z !\, b_z)^{-\frac{1}{2}}$
is a normalization factor and $H_{n_z}(\xi)$  are the Hermite polynomials. 
%The $\phi$-dependent part is given by
%\begin{equation}
%\psi_{\Lambda}(\phi)=\frac{1}{\sqrt{2\pi}}e^{i\Lambda \phi}.
%\end{equation} 

The normalized wave functions of the three-dimensional isotropic harmonic oscillator in spherical polar coordinates $(r, \theta, \phi)$ are given:
\begin{eqnarray}
|Nl\Lambda\rangle=\psi_{n_{r}l }(r)~ Y_{l \Lambda}(\theta,\phi),
\end{eqnarray} 
%where $N$ is the spherical principal quantum number which is defined as $N=2n_r+l$, 
where $n_r$ is the radial quantum number, $Y_{l \Lambda}$ is the spherical harmonic, 
$l$ is the orbital angular momentum. 
The radial part $\psi_{n_{r}l}$ is written in terms  of a dimensionless coordinate $\nu$ 
\begin{equation}
\psi^{l}_{n_r}(r)= C_{n_{r}l} \nu^{l/2}e^{-\frac{\nu}{2}}L_{n_{r}}^{(l+\frac{1}{2})}(\nu) 
\end{equation} 
with $\nu=\frac{r^2}{{b_0}^{2}}$,  where $b_0$ is the spherical oscillator %frequency
length,   
$C_{n_{r}l}= \bigg(\frac{2 n_r !}{(n_{r} +l +\frac{1}{2})!\, b_0^{3}}\bigg )^\frac{1}{2}$
is a normalization factor and $L^{(l+\frac{1}{2})}_{n_r}(\nu^2)$  are the associated Laguerre polynomials. 

The wave functions $|N n_z \Lambda\rangle$ of the deformed harmonic oscillator
can be decomposed over the spherical ones $|Nl\Lambda\rangle$ as 
\begin{equation}
|N n_z \Lambda\rangle= \sum_{n_r l \Lambda} A^{Nl}_{N n_z \Lambda}|Nl\Lambda\rangle
\end{equation} 
where $A^{Nl}_{N n_z \Lambda}=\langle N l \Lambda|N n_z \Lambda\rangle$ is the spatial overlap integral which can be numerically calculated in the spherical coordinate system as follows:
%polar ones using $\rho= r \sin\theta $ and $z = r \cos\theta$ and substitute with the wave functions, 
%the spatial overlap integral can be written as 
%\begin{eqnarray} 
%A^{Nl}_{N n_z \Lambda}&=&\sqrt{\frac{\pi^\frac{1}{2}}{2^{n_z} b_0^{\frac{9}{2}} b_\perp \sqrt{b_z}  n_z !}}~ 
% \int_{0}^\infty \int_{0}^\pi    r ^2
% e^{-\frac{r ^2}{2}(\frac{1}{b_{0}^{2}}+\frac {sin^2\theta}{b_{0} b_\perp} +  \frac {cos^2\theta}{b_{0} b_z})}~ 
%\nonumber\\
%&& R_{n_r l}(\frac{r}{b_0} )
%~R^{|\Lambda|}_{n_\rho}(  \frac{r sin \theta}{\sqrt{b_0 b_\perp}})~H_{n_z}( \frac{r cos \theta}{\sqrt{b_0 b_z}} )~ Y_{l \Lambda}(\theta,\phi=0)
%\nonumber\\
%&&sin \theta~ d r~ d \theta
%\end{eqnarray} 
%where $~~~~R^{m}_{n}(y)= \sqrt\frac{2 n !}{(n +m )!}~ y^m~ L^{m}_{n}(y^2)$.

\begin{eqnarray} 
A^{Nl}_{N n_z \Lambda}&=&\sqrt{2\pi}~ \int_{0}^\infty   \left(\int_{0}^\pi   
\psi^{|\Lambda|}_{n_\rho}(r \sin\theta)~\psi_{n_z}(r \cos\theta)
Y_{l \Lambda}^*(\theta,\phi=0) \sin \theta~d \theta \right)\psi_{n_{r}l }(r) r^2 d r.
\nonumber
\end{eqnarray} 

\newpage

\begin{figure}
%\includegraphics[width=14cm,height=8.cm,angle=270]{fig1.eps}
%\includegraphics[width=14cm,height=8.cm,angle=270]{ge0p.ps}
%\centerline{\psfig{figure=gen.ps,width=14cm,height=18cm,angle=270}}
\centerline{\includegraphics[scale=0.75]{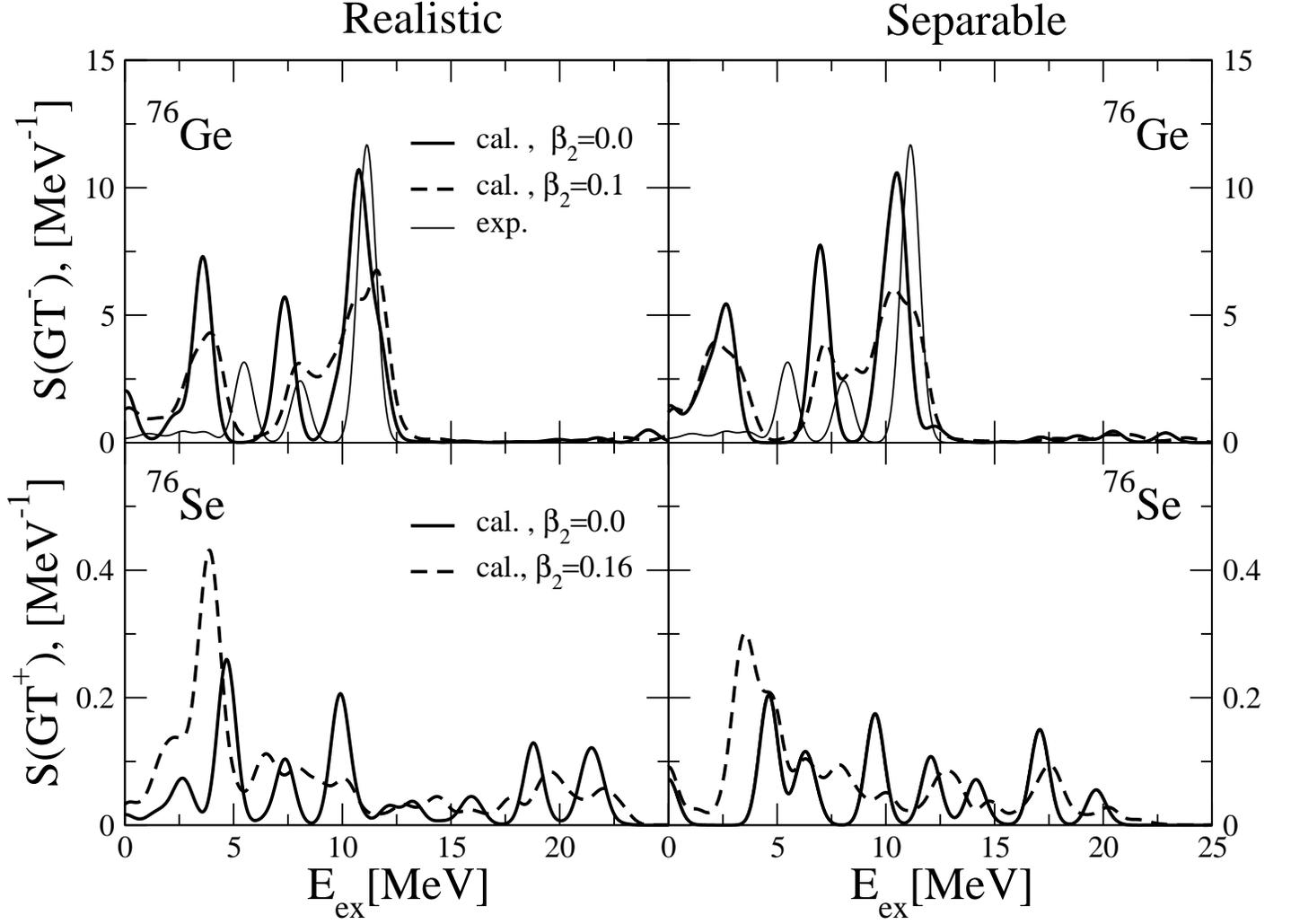}}
\caption{Gamow-Teller strength distributions $S{(GT^-)}$  in $^{76}$Ge and $S{(GT^+)}$  in $^{76}$Se 
as functions of the excitation energy $E_{ex}$ in the intermediate (for $^{76}$Ge$\rightarrow ^{76}$Se decay) 
nucleus $^{76}$As.
The QRPA calculation results obtained with realistic and separable forces 
are shown in panels (a) and (b), respectively.
Upper panels: 
the results corresponding to the spherical ($\beta_2=0.0$) and deformed ($\beta_2=0.1$) ground state of $^{76}$Ge 
are represented by the thick solid and dashed lines, respectively. 
Experimental data (thin solid line) are from~\cite{madey}. 
Lower panels: The results corresponding to the spherical ($\beta_2=0.0$) and deformed ($\beta_2=0.16$) ground state of $^{76}$Se are represented by the thick solid and dashed lines, respectively. 
}
\label{fig.1}
\end{figure}

\begin{figure}
\centerline{\includegraphics[scale=0.75]{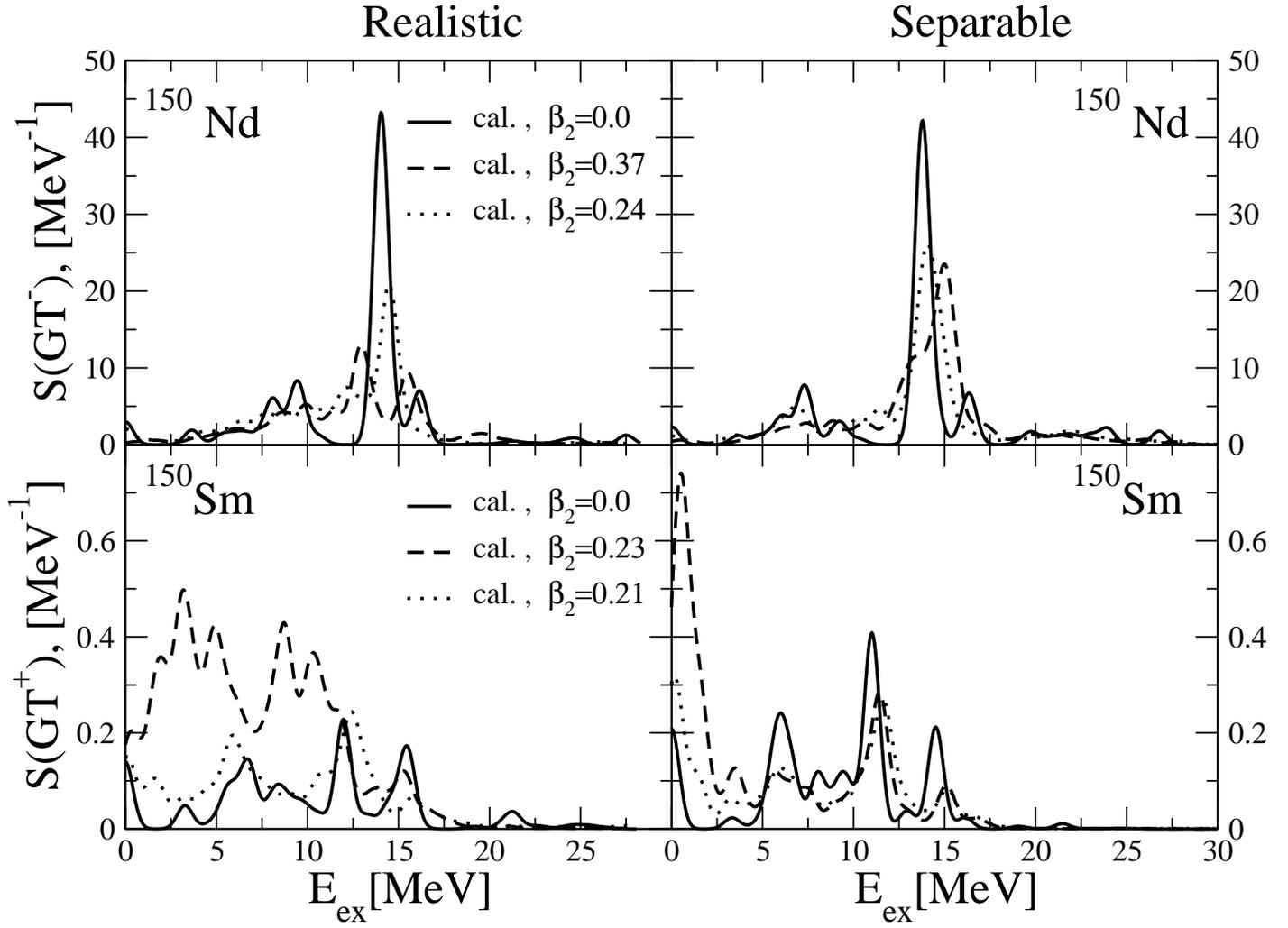}}
\caption{The same as in Fig. 1, but for $^{150}$Nd and $^{150}$Sm.} 
\label{fig.5}
\end{figure}
 
\begin{figure}
\centerline{\includegraphics[scale=0.7]{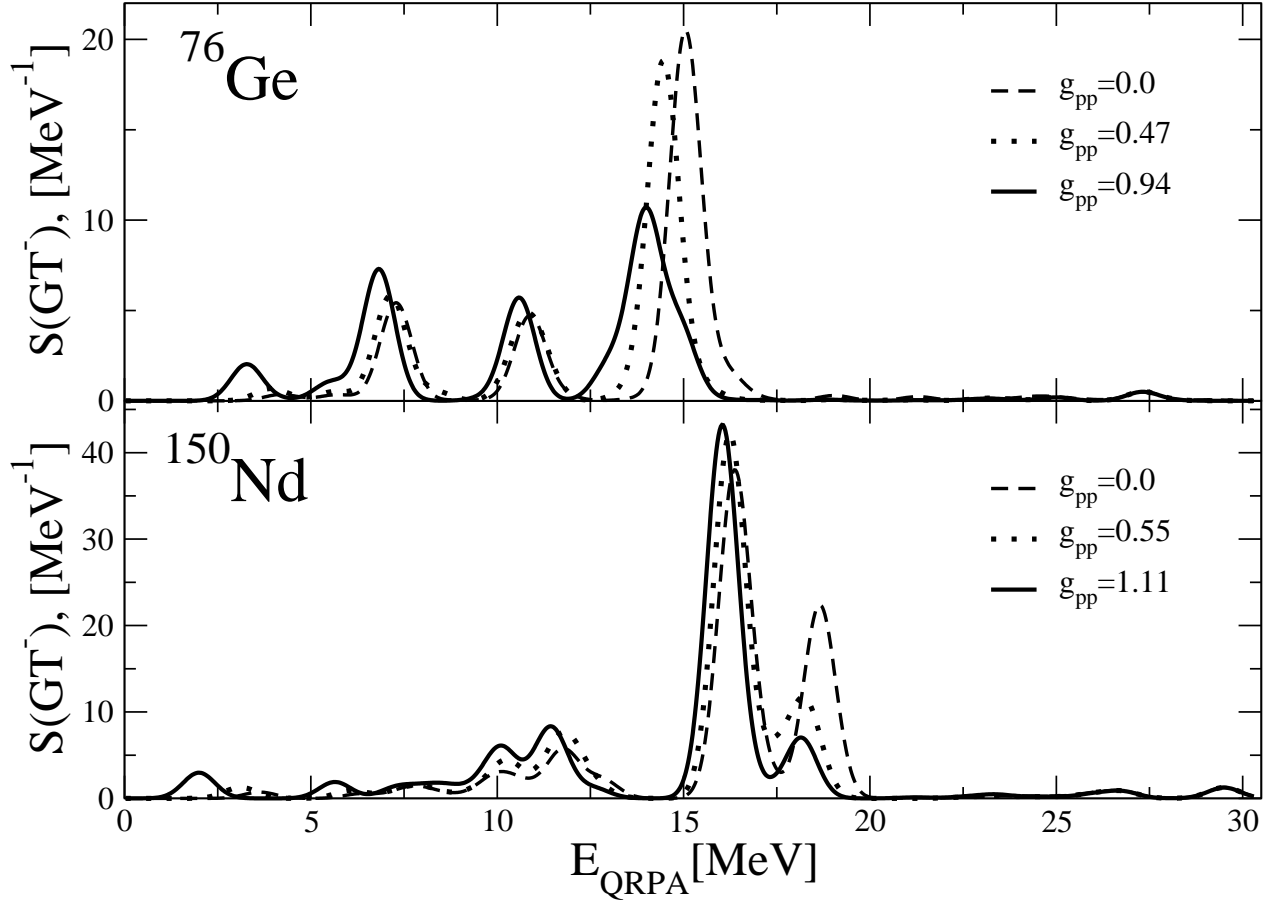}}
\caption{Gamow-Teller strength distributions $S{(GT^-)}$  in $^{76}$Ge and $^{150}$Nd (spherical limit) for different 
particle--particle interaction strengths $g_{pp}$ %for $\beta_2=0.0$ 
as functions of the QRPA energy $E_{QRPA}$. The solid lines (the same as in Fig.~1,2, upper left panels) represent $S{(GT^-)}$ corresponding to the fitted $g_{pp}$ (Table I). The dashed and the dotted lines correspond to smaller $g_{pp}$ as indicated in the Figure.} 
\label{fig.55}
\end{figure}
 
\begin{figure}\centerline{\includegraphics[scale=0.75,angle=-90]{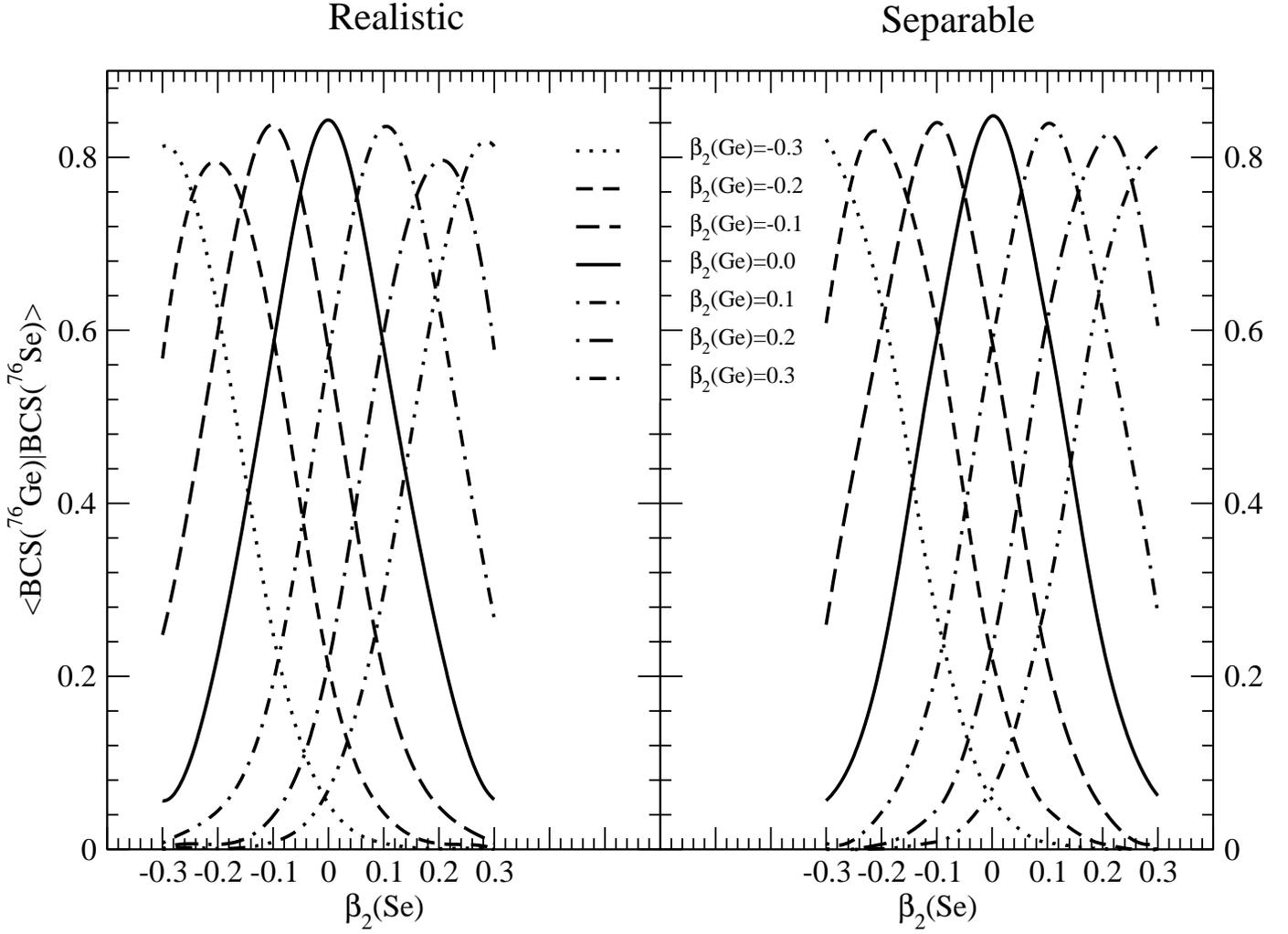}}
\caption{The dependence of the BCS overlap factor in Eq.~(\ref{8}) on the difference in deformations of $^{76}$Ge 
and $^{76}$Se. The calculation results obtained with realistic and separable forces are shown in panels (a) and (b), respectively.}
\label{fig.11}
\end{figure}

\begin{figure}
\centerline{\includegraphics[scale=0.75]{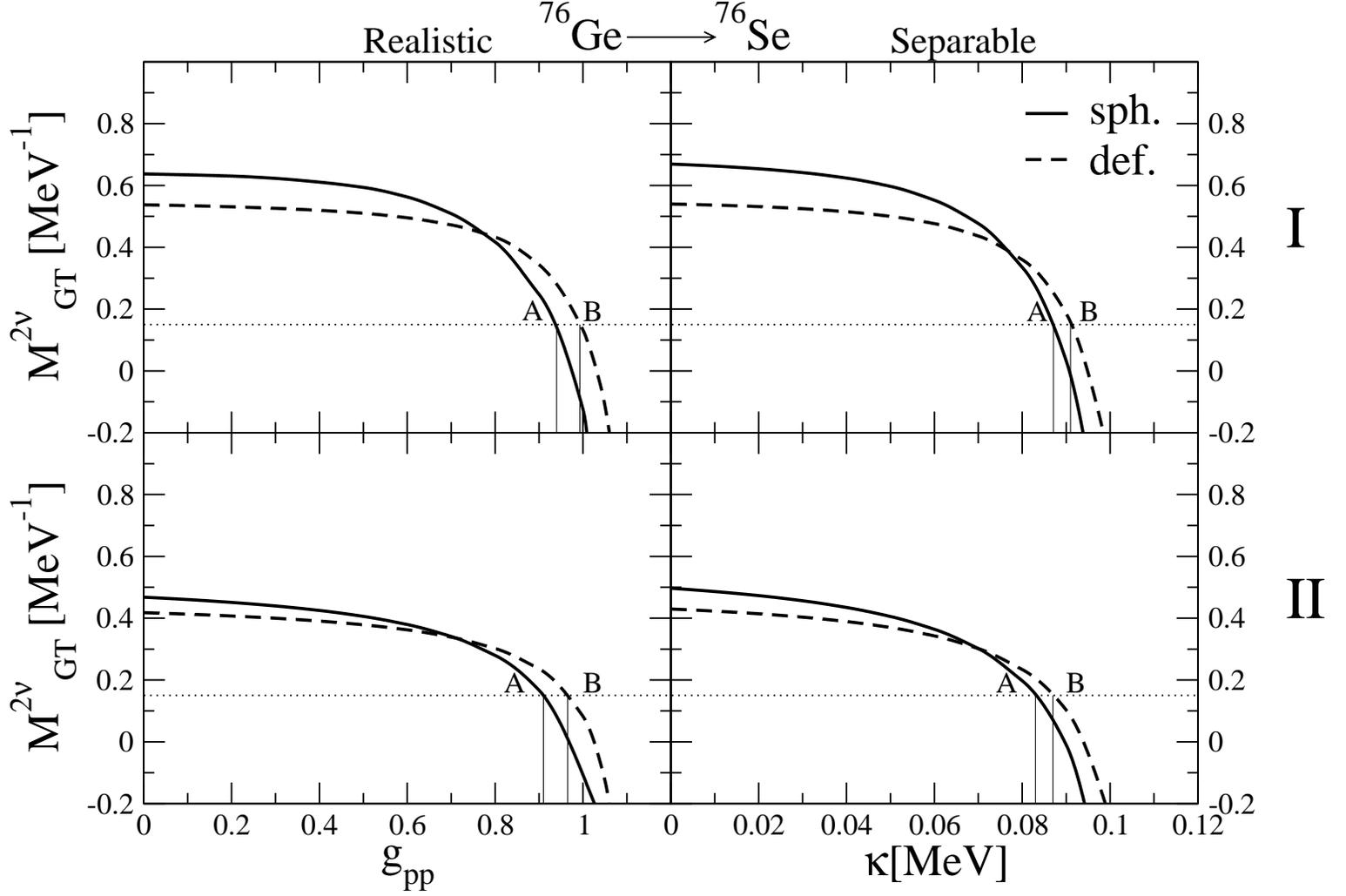}}
\caption{$2\nu\beta\beta$-decay matrix element for $^{76}$Ge$\to^{76}$Se decay as function of 
particle--particle interaction strength $g_{pp}$ of the realistic forces (left panels), 
and of $\kappa$ of the separable forces (right panels). 
The solid line (``sph.") corresponds to spherical shape of the initial and final nuclei. 
The dashed one (``def.") is associated with the  deformation parameters taken from Ref.~\cite{ragha} 
($\beta_2(^{76}$Ge)$=0.1$, $\beta_2(^{76}$Se)$=0.16$). The doted horizontal line
corresponds to experimental $M^{2\nu-exp}_{GT}$ obtained in Ref.~\cite{barab} using  $g_A=1.25$. 
The points A and B in each panel specify the values of the $p-p$ interaction for which the value of $M^{2\nu-exp}_{GT}$ 
is fitted. The upper (case I) and lower (case II) panels correspond to the calculations with the shifted and 
unshifted QRPA spectrum, respectively, as explained in the text.}
\label{fig.7}
\end{figure}

\begin{figure}
\centerline{\includegraphics[scale=0.75]{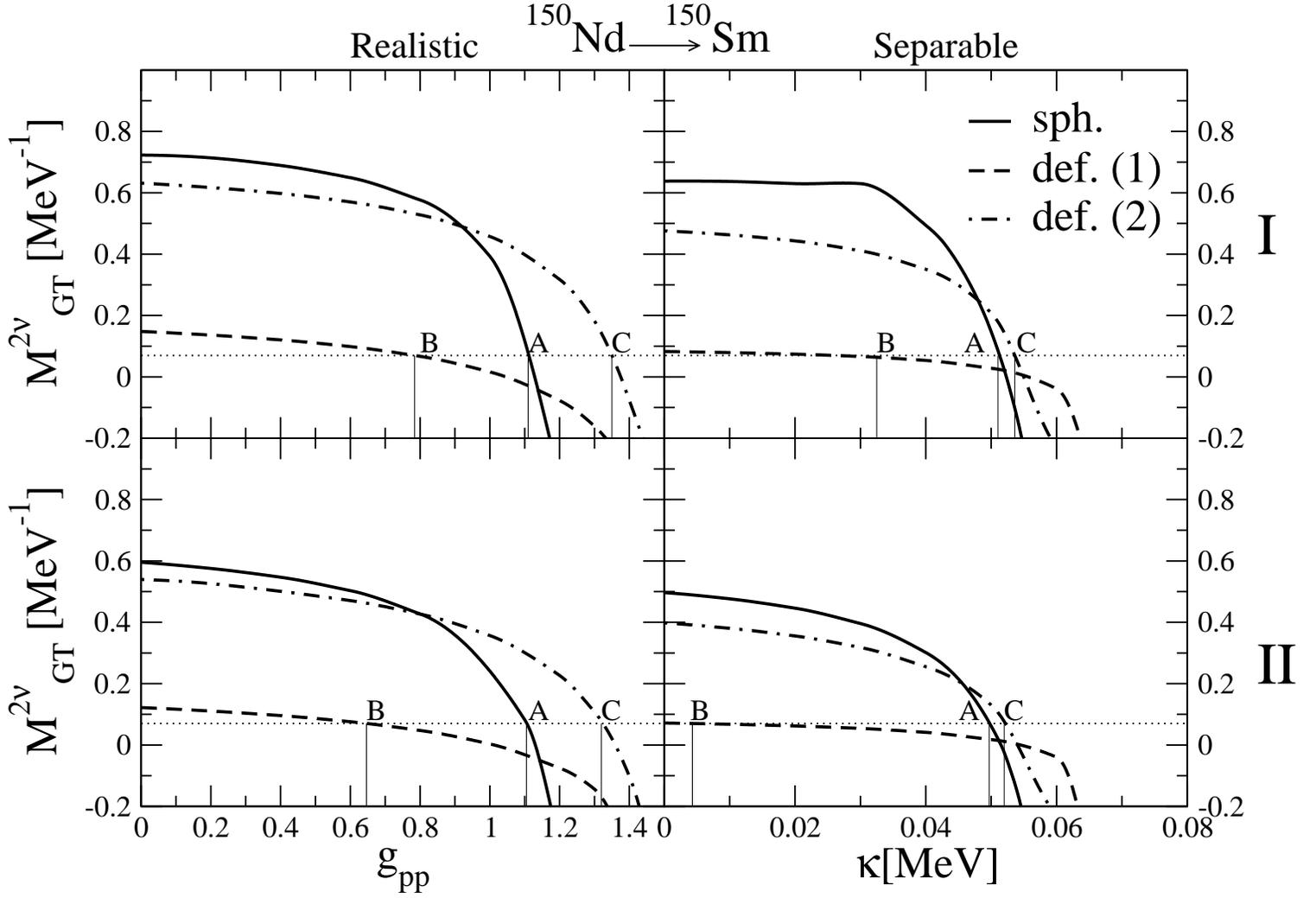}}
\caption{
The same as in Fig. 5, but for $^{150}$Nd$\to^{150}$Sm decay with 2 different sets of deformation parameters: from Ref.~\cite{ragha}
($\beta_2(^{150}$Nd)$=0.37$, $\beta_2(^{150}$Sm)$=0.23$, ``def. (1)" ) and from Ref.~\cite{moeller}
($\beta_2(^{150}$Nd)$=0.24$, $\beta_2(^{150}$Sm)$=0.21$, ``def. (2)").}
\label{fig.9}
\end{figure}

\end{document}